\documentclass[runningheads]{llncs}

\AtBeginDocument{
  \setlength\abovedisplayskip{2pt}
  \setlength\belowdisplayskip{2pt}}

\usepackage{cite}
\usepackage{amsmath,amssymb,amsfonts}
\usepackage[inline]{enumitem}
\usepackage{graphicx}
\usepackage{textcomp}
\usepackage{xcolor}
\usepackage{algorithmic}
\usepackage{algorithm}
\usepackage{booktabs}
\usepackage{subcaption}
\usepackage{caption}
\usepackage{tabularx}
\usepackage{hyperref}
\usepackage{url} 
\usepackage{gensymb}

\begin{document}

\title{Characterizing GPU Energy Usage in Exascale-Ready Portable Science Applications}

\titlerunning{GPU Energy Exascale Port Apps}

\author{
William F. Godoy\inst{1}\orcidID{0000-0002-2590-5178} 
\and
Oscar Hernandez\inst{1}\orcidID{0000-0002-5380-6951} 
\and
Paul R. C. Kent\inst{1}\orcidID{0000-0001-5539-4017}
\and
Maria Patrou\inst{1}\orcidID{0000-0003-3975-4638} 
\and \\
Kazi Asifuzzaman\inst{1}\orcidID{0000-0002-4004-4791}
\and
Narasinga Rao Miniskar\inst{1}\orcidID{0000-0001-8259-8891}
\and
Pedro Valero-Lara\inst{1}\orcidID{0000-0002-1479-4310}
\and
Jeffrey S. Vetter\inst{1}\orcidID{0000-0002-2449-6720}
\and
Matthew D. Sinclair\inst{2}\orcidID{0000-0003-0189-7895}
\and
Jason Lowe-Power\inst{3}\orcidID{0000-0002-8880-8703}
\and
Bobby R. Bruce\inst{3}\orcidID{0000-0001-6070-9722}
}

\authorrunning{Godoy et al.}
%
\institute{
Oak Ridge National Laboratory, Oak Ridge, TN, USA\\ 
\email{\{godoywf,oscar,kentpr,patroum,asifuzzamank,miniskarnr,valerolarap,vetter\}@ornl.gov}
\and
University of Wisconsin-Madison, Madison, WI, USA\\ 
\email{\{sinclair\}@cs.wisc.edu} 
\and
University of California, Davis, Davis, CA, USA\\
\email{\{jlowepower,bbruce\}@ucdavis.edu}
}

\maketitle

\vspace{-0.7cm}
\begin{abstract}
We characterize the GPU energy usage of two widely adopted exascale-ready applications representing two classes of particle and mesh solvers: (i)~QMCPACK, a quantum Monte Carlo package, and (ii)~AMReX-Castro, an adaptive mesh astrophysical code.
We analyze power, temperature, utilization, and energy traces from double-/single (mixed)-precision benchmarks on NVIDIA's A100 and H100 and AMD's MI250X GPUs using queries in NVML and rocm\_smi\_lib, respectively.
We explore application-specific metrics to provide insights on energy vs. performance trade-offs.
Our results suggest that mixed-precision energy savings range between 6-25\% on QMCPACK and 45\% on AMReX-Castro.
Also, we found gaps in the AMD tooling used on Frontier GPUs that need to be understood, while query resolutions on NVML have little variability between 1\,ms-1\,s. 
Overall, application level knowledge is crucial to define energy-cost/science-benefit opportunities for the codesign of future supercomputer architectures in the post-Moore era.

\end{abstract}

\vspace{-2ex}
\section{Introduction}
\label{sec:Introduction}
\vspace{-1ex}

As energy consumption and costs have grown exponentially in the post-Moore exascale era, high-performance computing (HPC) faces new challenges~\cite{reed2022reinventinghighperformancecomputing}.
Interest in the energy-cost/science-benefit trade-offs is again gaining traction\footnote{\url{https://www.orau.gov/2024EECWorkshop}} as HPC systems become more heterogeneous~\cite{osti_1473756}.
Since HPC traditionally focused on optimizing time-to-solution, it is crucial to understand applications characteristics to design future energy-efficient hardware.
Here, we provide insights on the GPU energy characteristics of two applications developed under the US Department of Energy's (DOE's) Exascale Computing Project (ECP, 2016--2023)~\cite{8528398} that are widely used at HPC production facilities across the world: (i) QMCPACK~\cite{kim2018qmcpack,doi:10.1063/5.0004860,Luo-qmcpack}, a quantum Monte Carlo (QMC) package, and (ii) the mesh-based AMReX-Castro's~\cite{almgren2020castro} compressible astrophysics code.
We capture power, utilization, temperature, and calculate energy traces on NVIDIA's A100 and H100 GPUs and AMD's MI250X GPU.
To capture these measurements, we designed an open-source \texttt{HWEnergyTracer.jl}\footnote{\url{https://github.com/JuliaORNL/HWEnergyTracer.jl}} tool that runs side-by-side with an application and captures queries from NVIDIA's Management Library (NVML), and AMD's ROCm System Management Interface Library (\texttt{rocm\_smi\_lib}).

The paper is organized as follows: Section~\ref{sec:Background} provides information for QMCPACK, AMReX-Castro and selected benchmarks. 
Section~\ref{sec:Methodology} describes our methodology and the targeted GPU systems.
Section~\ref{sec:Results} presents our results and analysis of the applications' energy characteristics.
Related work in HPC is shown in Section~\ref{sec:Related-Work}. Finally, Section~\ref{sec:Conclusions} provides our conclusions and future directions. To the best of our knowledge, our contributions on quantifying science-per-energy has not previously been an integral part of the applications' development process.

\vspace{-2ex}
\section{Background}
\label{sec:Background}
\vspace{-1.5ex}

\paragraph{QMCPACK and the NiO benchmark:} 
QMCPACK is an open-source, many-body, ab-initio QMC framework solving the Schr\"odinger equation for atoms, molecules, 2D nanomaterials, and solids.
QMC methods lead to far greater accuracy, but at a much greater computational cost.
Key recent QMCPACK improvements made during the ECP included (i) a redesigned diffusion Monte Carlo (DMC) solver~\cite{Luo-qmcpack} -- the focus on this study -- using OpenMP offload capabilities on GPUs, and (ii) software engineering improvements for CPU/GPUs~\cite{GODOY2025107502}.
We use the nickel oxide (NiO) supercell benchmark~\footnote{\url{https://www.olcf.ornl.gov/wp-content/uploads/OLCF-6_QMCPACK_description-1.pdf}} which is characterized
by the total number of electrons, for which its required memory grows quadratically, while its computation increases at a cubic rate.

\paragraph{AMReX-Castro and the Sedov case:} 
AMReX~\cite{AMReX_JOSS} is a widely used adaptive mesh refinement framework that powers several HPC applications running at scale.
AMReX decomposes a problem into levels of adaptive resolution and rectangular patches.
During the ECP, AMReX's solver capabilities were advanced for new CPU/GPU architectures, while energy-efficiency is in the product roadmap~\cite{doi:10.1177/10943420241271017}.
We use the Castro astrophysical radiation-magneto-hydrodynamics code that builds on AMReX to run the 2D Sedov spherical blast wave standard problem on a rectangular AMR mesh.
Its computational costs are typically driven by the time evolution of the total number of AMR cells~\cite{9835553}.

\section{Methodology}
\label{sec:Methodology}

We use the Julia language~\cite{Bezanson2017-ca} to call queries listed in Table~\ref{tab:queries} in NVIDIA's NVML by using the CUDA.jl package~\cite{10.1109/TPDS.2018.2872064}, and AMD's \texttt{rocm\_smi\_lib} C API using Julia's foreign function interface.
The result is the \texttt{HWEnergyTracer.jl} tool
that runs 15\,s before the start and 10\,s after the end of the application run to guarantee a steady static power state.
Power values are integrated over time to obtain energy traces, and to minimize observed false positives (over-/under-shoots). 
We use three different GPU systems, 2 NVIDIA and 1 AMD, as described in Table~\ref{tab:systems}.
No power capping was applied in this work.

\begin{table}[tb!]
    \footnotesize
    \centering
    \caption{Queries used for NVIDIA's NVML and AMD's rocm\_smi\_lib}
    \vspace{-1ex}
    \begin{center}
        \begin{tabular}{|p{0.2\linewidth}|p{0.22\linewidth}|p{0.63\linewidth}|} 
         \hline
         \textbf{Metric} & \textbf{Relevant Query} & \textbf{Description} \\
         \hline
         \hline
         \textbf{NVIDIA} & \textbf{nvmlDeviceGet*} &                      \\
         \hline
         Power (W) & PowerUsage  &  Power usage of the GPU and its associated circuitry (e.g., memory) averaged over a 1~s interval~\cite{YangAdamek2024-nvidiaSMIAccuracy} \\
         \hline
         Utilization (\%) &  UtilizationRates  &  Percent of time over the past sample period, between 1 and $\scriptstyle\frac{1}{6}$~s, during which kernels were executing \\
         \hline
         Temperature (\degree C)  & Temperature   &  Current temperature readings for the device \\
         \hline
         \textbf{AMD} & \textbf{rsmi\_dev\_*}   &                      \\
         \hline
         Power (W) & power\_ave\_get  &  device energy counter average for a short time (1\,ms) \\
          \hline
         Utilization (\%) &  busy\_percent\_get  &  Percentage of time busy processing \\
         \hline
         Temperature (\degree C)  &  temp\_metric\_get  &  Retrieved from the temperature sensor for the device \\
          \hline
        \end{tabular}
    \end{center}
    \label{tab:queries}
    \vspace{-2ex}
\end{table}

\begin{table}[tb!]
  \vspace{-3ex}
  \footnotesize
  \caption{System hardware and software used in this study}
  \vspace{-1ex}
  \footnotesize{
    \begin{center}
      \begin{tabular}{ |l|c|c|c| } 
       \hline
       \textbf{System} & \textbf{Milan0} & \textbf{Hudson} & \textbf{Frontier} \\
       \hline
       \hline
       \textbf{Hardware} &            &             &            \\
       GPU-per-node       & 2 NVIDIA A100 & 2 NVIDIA H100 & 8 GCD AMD MI250X \\
       Memory(GB)/Bandwidth(GB/s)           & HBM2E 80/1,940      & HBM3 94/1,940 & HBM2E 64/3,276 \\
       Thermal Design Power (W)  & 300   &    400   &  500  \\
       \hline
       \hline
       \textbf{Software} &  & & \\
       GPU Tool Chain  & NVHPC 24.9 & NVHPC 24.9 & ROCm 6.2 \\ 
       \hline 
       QMCPACK & v3.17.1 & v3.17.1 & v3.17.1 \\
       Compiler  & Clang 19.1 & Clang 19.1 & AMDClang 6.2\\
       Programming Model & OpenMP-offload & OpenMP-offload & OpenMP-offload \\
       \hline
       AMReX-Castro & v24.12 & v24.12 & v24.12 \\
       Compiler & GCC 13.2     & GCC 13.2     & GCC 12.3 \\
       Programming Model & CUDA 12.4 & CUDA 12.4 & HIP 6.2 \\
      \hline
    \end{tabular}
  \end{center}
}
\label{tab:systems}
\vspace{-2ex}
\end{table}

\vspace{-2ex}
\section{Results}
\label{sec:Results}
\vspace{-2ex}

Results are presented for the application energy characteristics traces. We discuss: (i) regions of interests based on power consumption, (ii) measurement resolutions from 1\,ms to 1\,s to capture difference of the averages values given by the vendor tools, (iii) impact of double/single precision, and (iv) exploring application-specific energy-efficiency metrics (e.g. science/Joule).

\subsection{QMCPACK NiO benchmark results}

\begin{figure}[tb!]
\centering
\subfloat[5 steps x 3 blocks = 15 iterations]{\includegraphics[width=0.5\linewidth,height=0.27\linewidth]{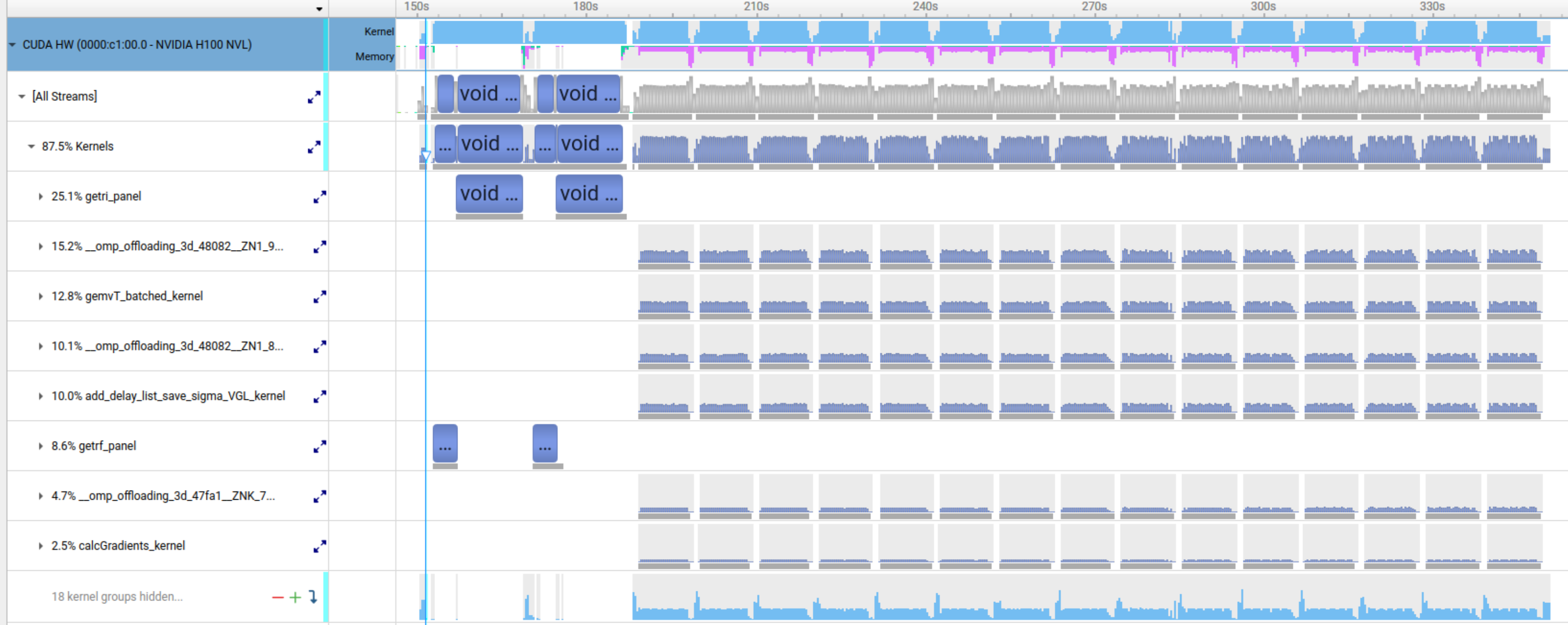}
\label{fig:trace_qmcpack_dmc}
}%
\subfloat[1\,ms snapshot]{\includegraphics[width=0.5\linewidth,height=0.27\linewidth]{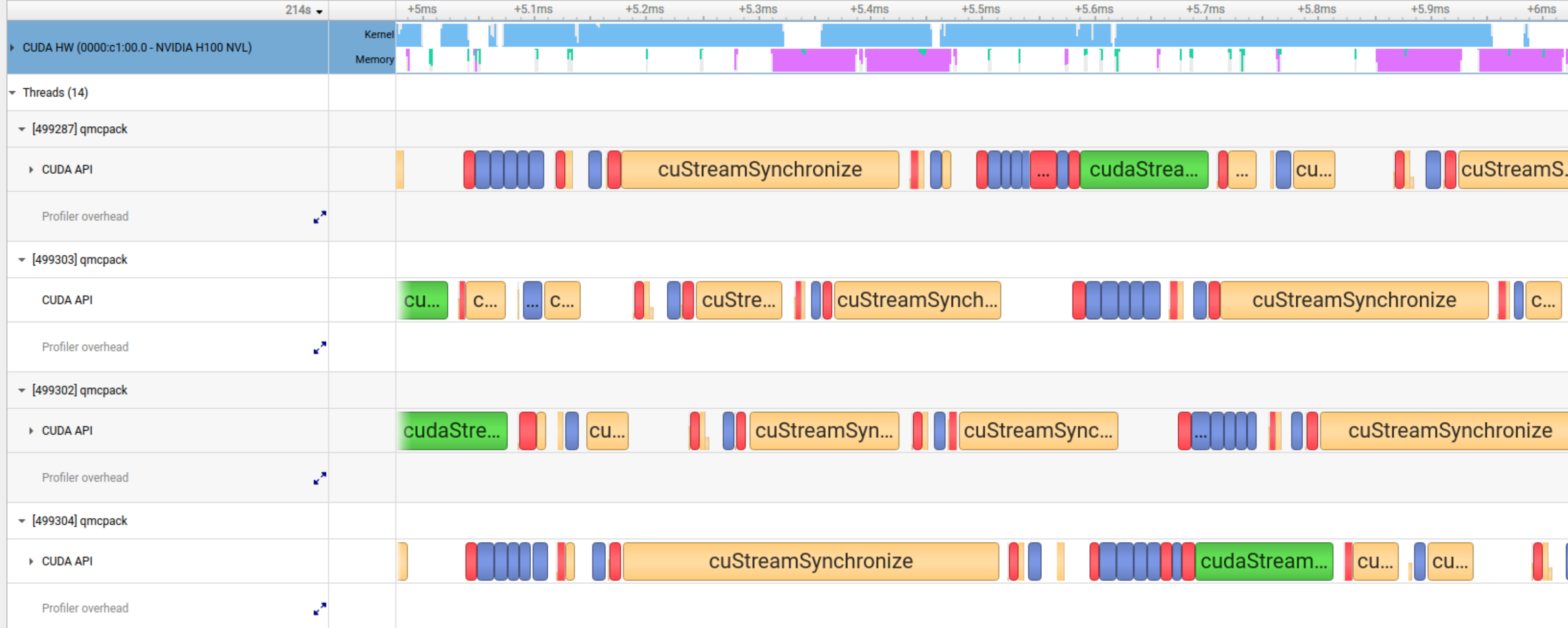}
\label{fig:trace_qmcpack_dmc_1ms}
}
\caption{QMCPACK NiO Benchmark DMC GPU traces on an NVIDIA H100.}
\label{fig:trace_qmcpack}
\end{figure}

We run the NiO benchmark for a representative supercell of 512 atoms and 6,144 electrons. Fig.~\ref{fig:trace_qmcpack} shows the DMC code profile showing the asynchronous nature of kernel launching for each MC ``walker'' to maximize GPU usage (blue top row)~\cite{Luo-qmcpack}. The number of walkers is maximized to fit on GPU memory~\cite{10.1145/3581576.3581621}.

\begin{figure}[tb!]
  \centering
  \subfloat[H100 1\,ms]{\includegraphics[width=0.3\textwidth,height=0.2\textwidth]{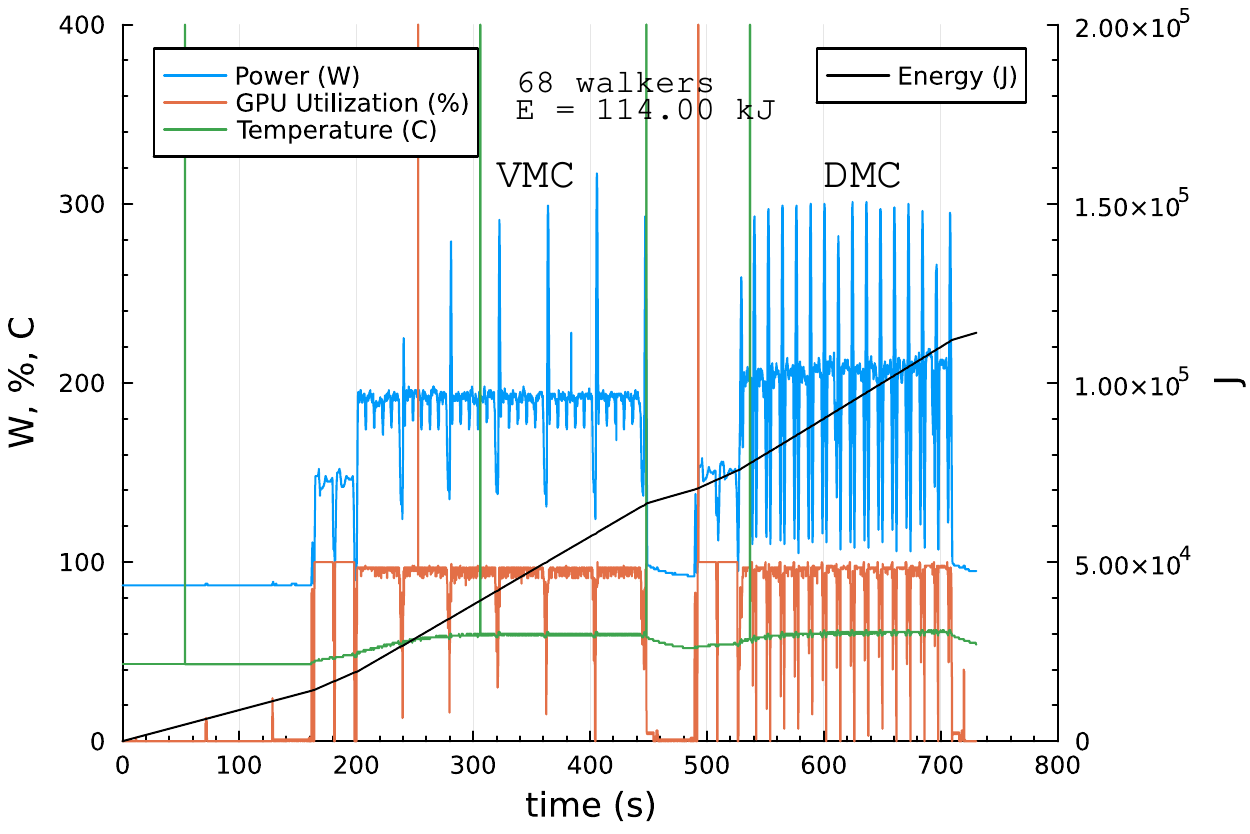}\label{fig:resolution_qmcpack_1ms_h100}}%
  \subfloat[H100 10\,ms]{\includegraphics[width=0.4\textwidth,height=0.3\textwidth]{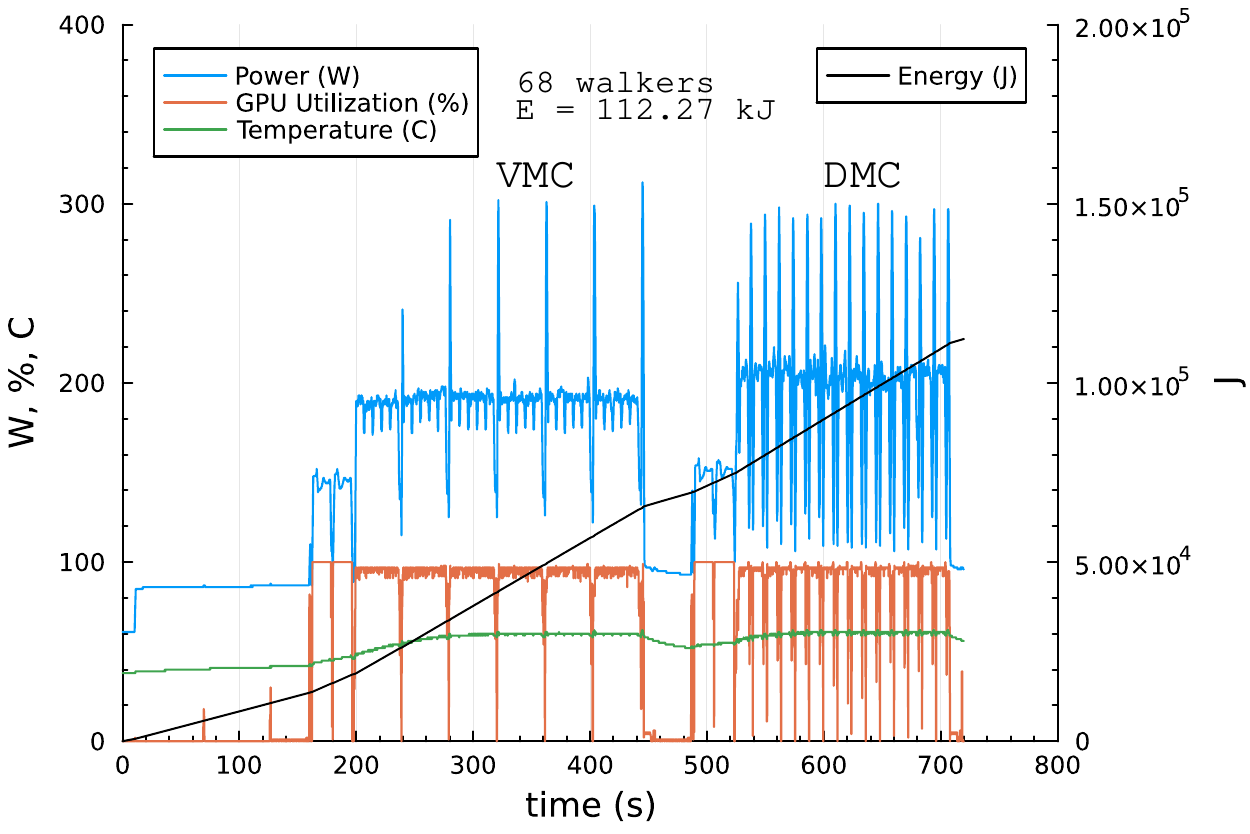}
  }%
  \subfloat[H100 1\,s]{\includegraphics[width=0.3\textwidth,height=0.2\textwidth]{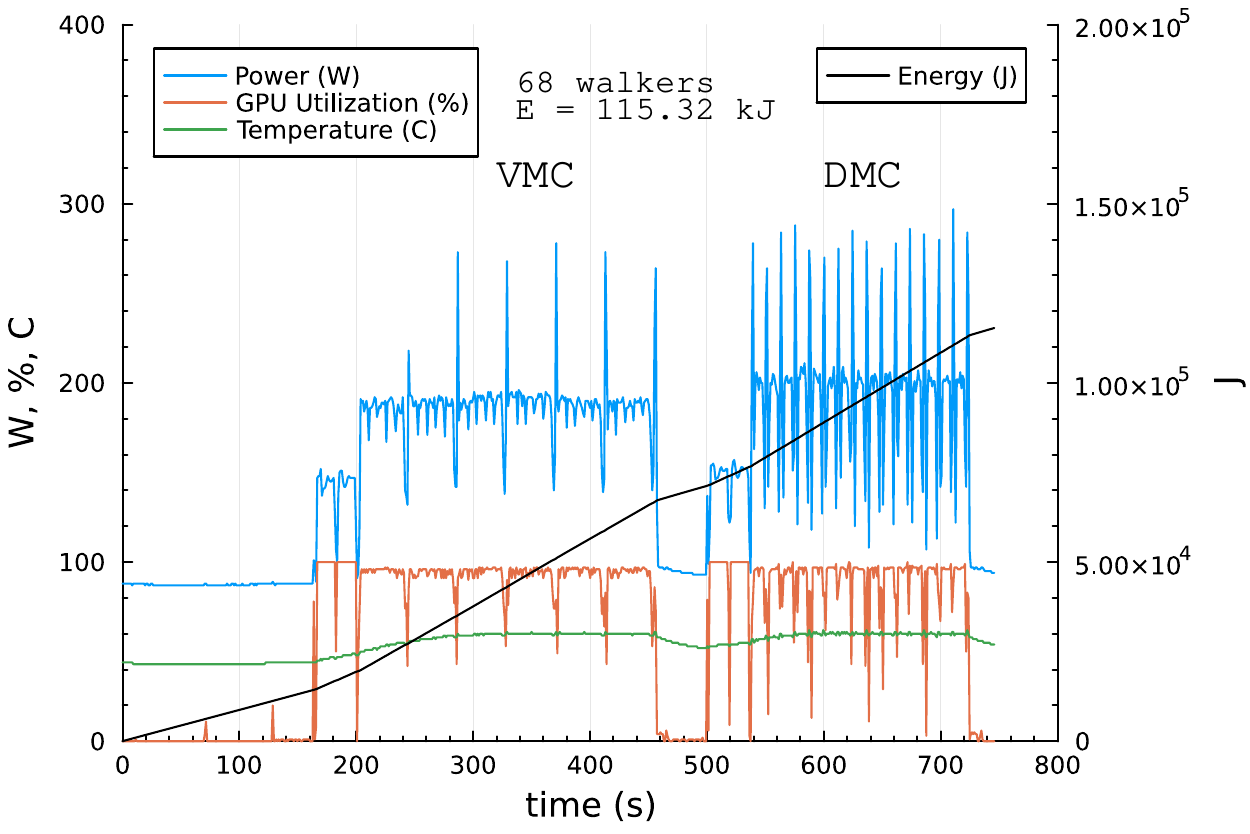}
  }
  
  \subfloat[A100 1\,ms]{\includegraphics[width=0.3\textwidth,height=0.2\textwidth]{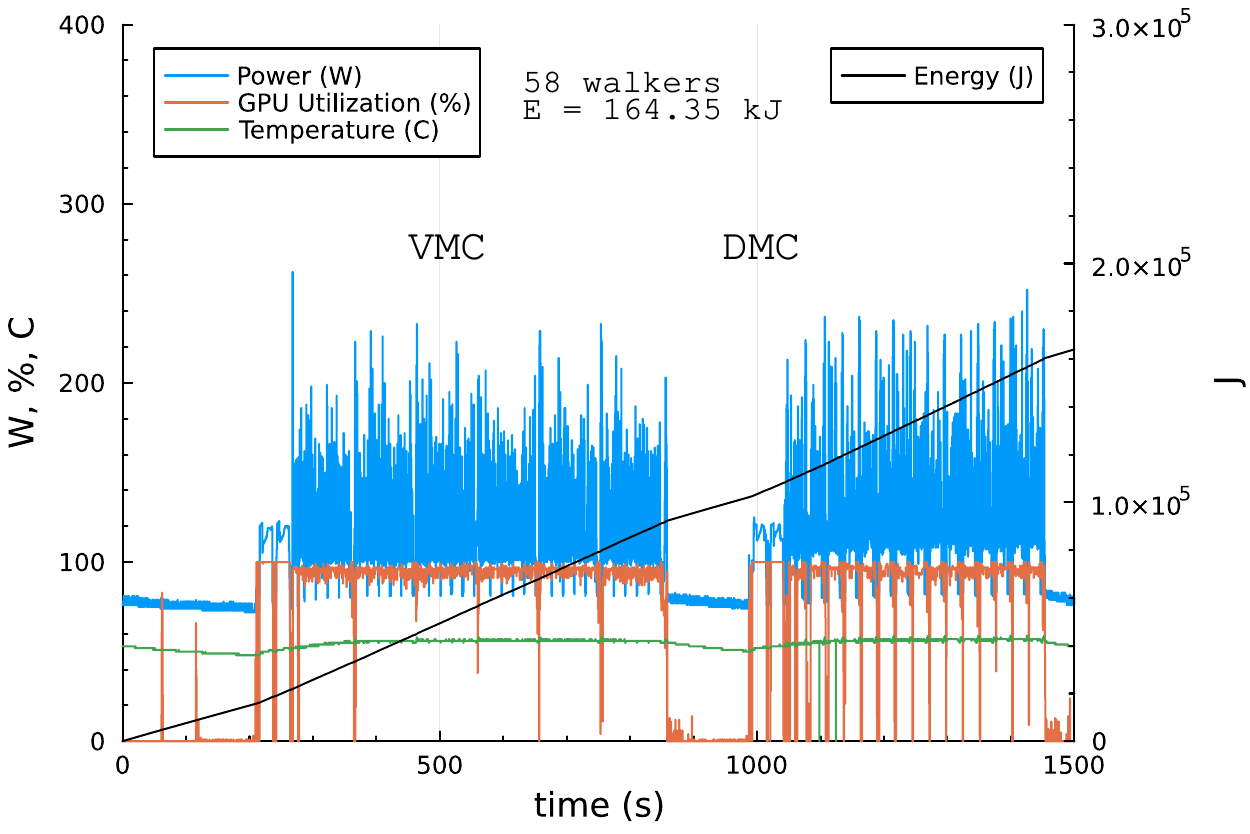}\label{fig:resolution_qmcpack_1ms_a100}}%
  \subfloat[A100 10\,ms]{\includegraphics[width=0.4\textwidth,height=0.3\textwidth]{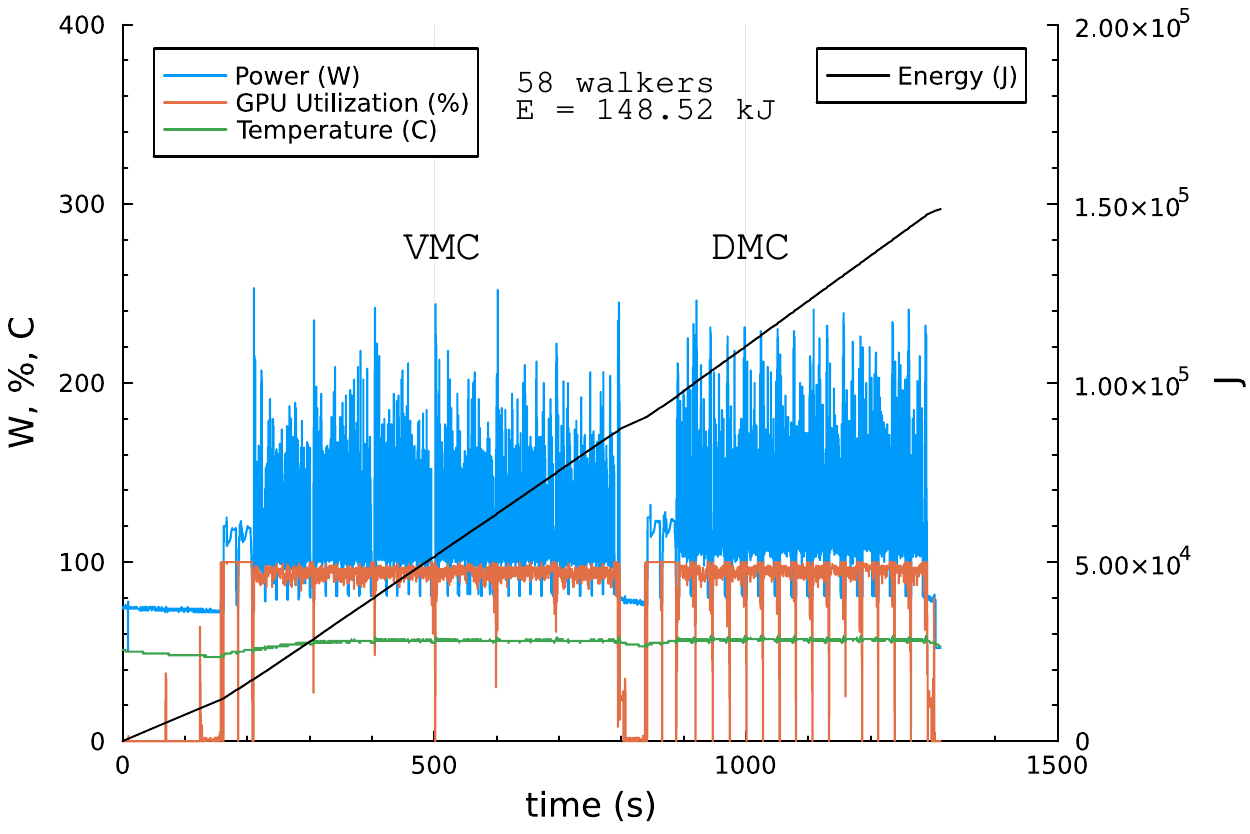}
  }%
  \subfloat[A100 1\,s]{\includegraphics[width=0.3\textwidth,height=0.2\textwidth]{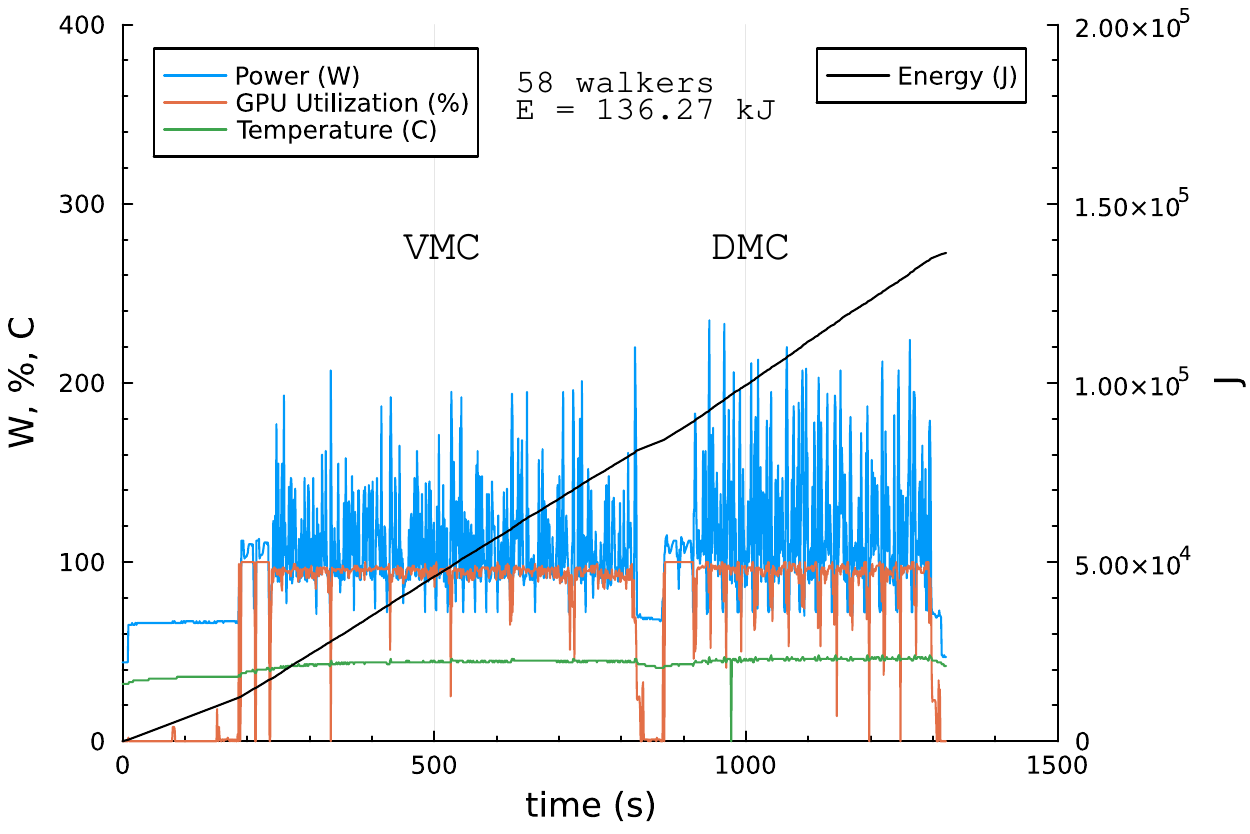}
  }
  \caption{Energy characteristics of the QMCPACK NiO benchmark on NVIDIA H100 and A100 for different query time resolutions.}
  \label{fig:resolution_qmcpack_nvidia}
  \vspace{-2ex}
\end{figure}

\begin{figure}[tb!]
  \centering
  \subfloat[MI250X 1\,ms]{\includegraphics[width=0.4\textwidth,height=0.3\textwidth]{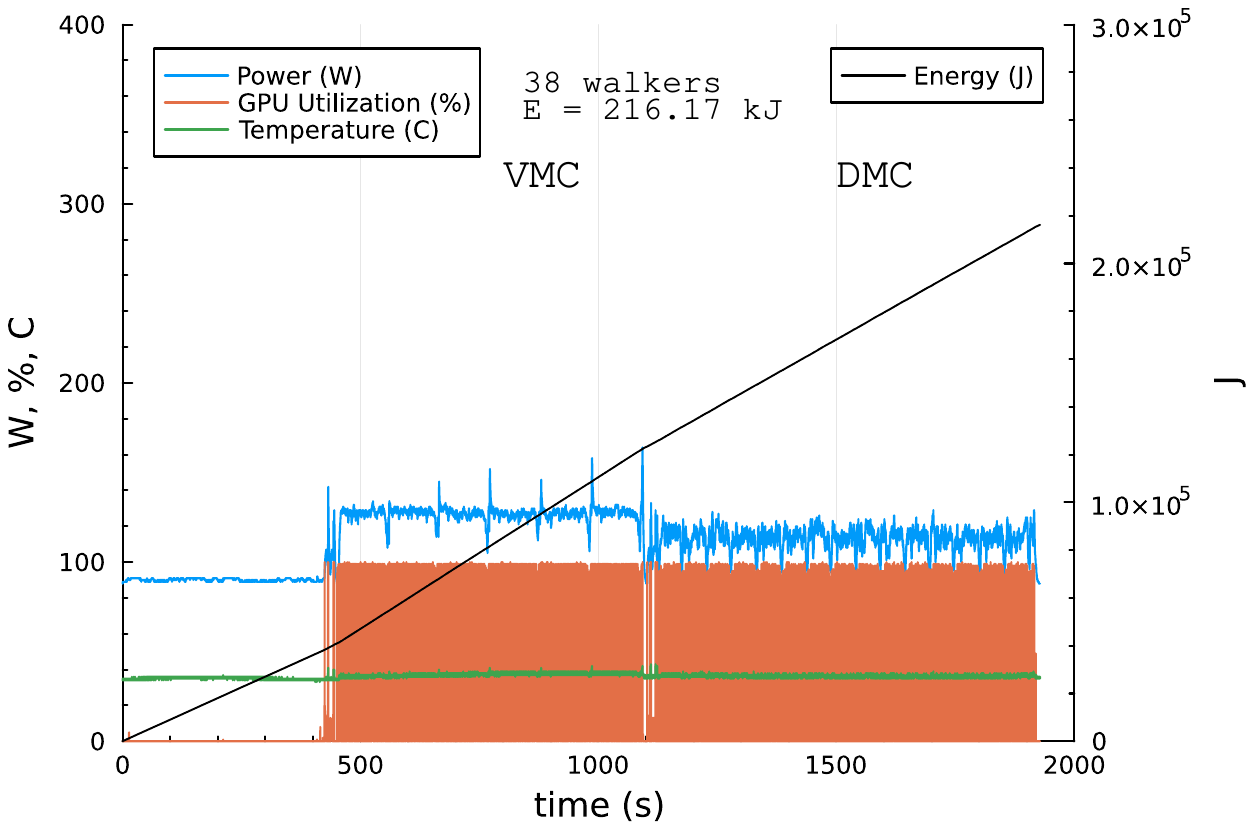}
  }%
  \subfloat[MI250X 10\,ms]{\includegraphics[width=0.3\textwidth,height=0.2\textwidth]{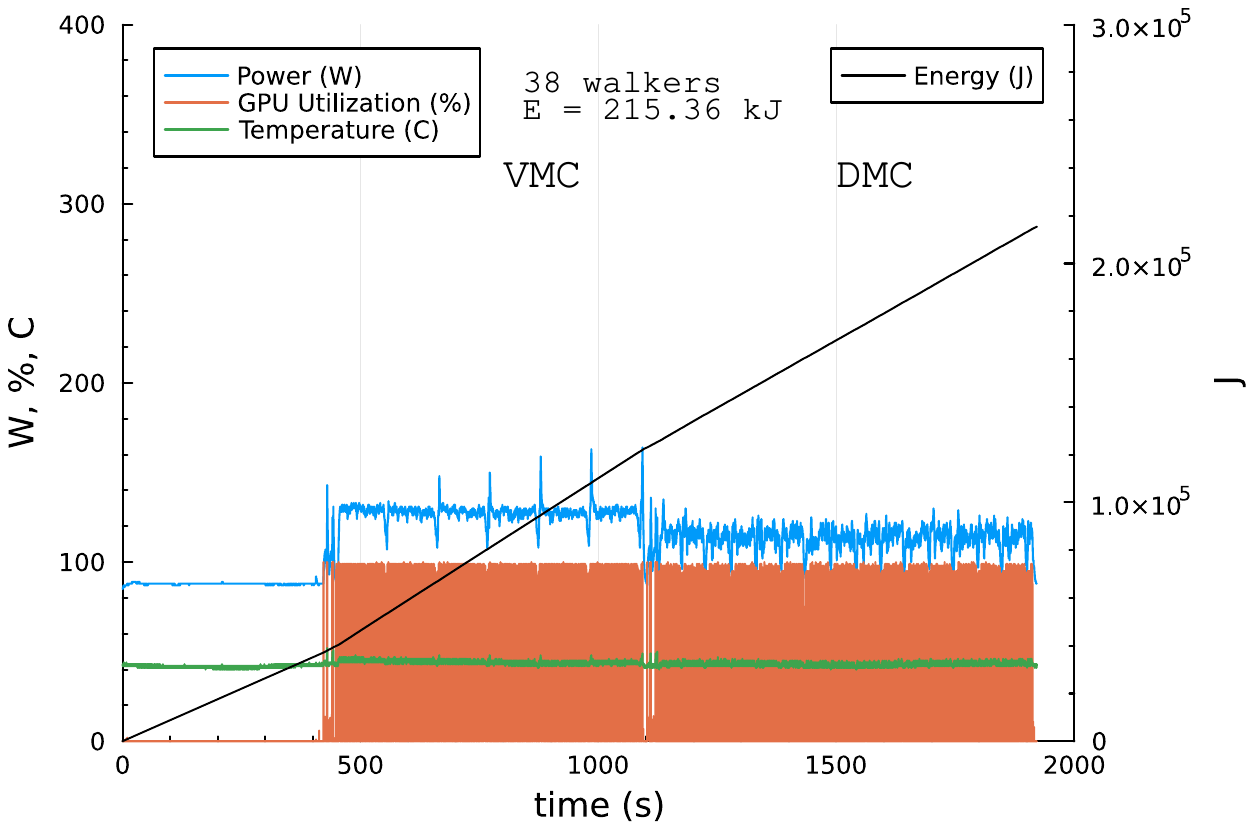}
  }%
  \subfloat[MI250X 1\,s]{\includegraphics[width=0.3\textwidth,height=0.2\textwidth]{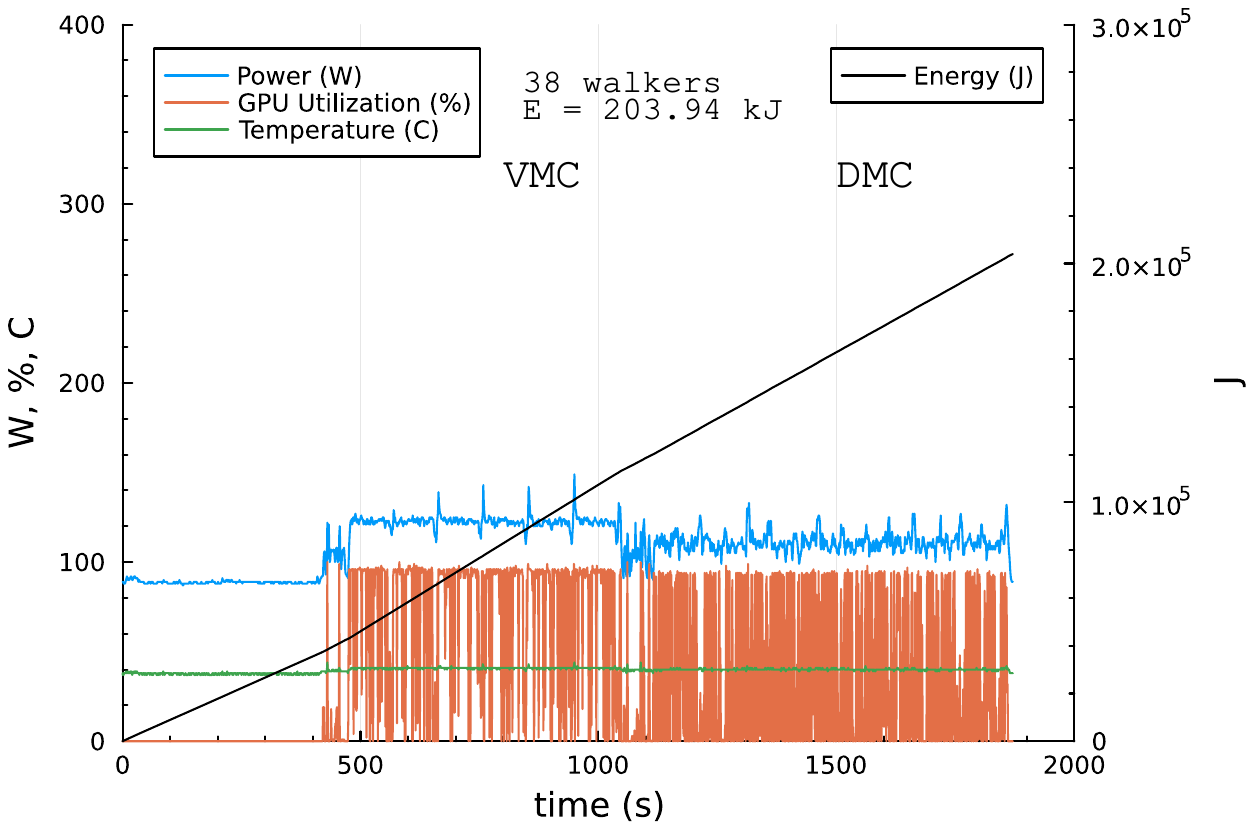}}

  \caption{Energy characteristics of the QMCPACK NiO benchmark on an AMD MI250X for different query time resolutions.}
  \label{fig:resolution_qmcpack_amd}
  \vspace{-2ex}
\end{figure}

Figs.~\ref{fig:resolution_qmcpack_nvidia} and ~\ref{fig:resolution_qmcpack_amd} show the power, temperature, utilization and energy traces for the maximum number of walkers on the NVIDIA H100 and A100, and AMD MI250X GPUs, respectively, for different query time resolutions. In all cases, we use double-precision configurations.
The query time resolution used in subsequent runs is highlighted as the larger figure.
Also, the four run stages are captured by the traces for spline data initialization (showing low GPU activity), Variational Monte Carlo (VMC), matrix inversions, and DMC.

On the H100 and A100, the DMC power and utilization patterns match the kernel behavior shown in Fig.~\ref{fig:trace_qmcpack_dmc}. As for the MI250X, the DMC stage power shows similar characteristics for minimum values, but peaks are not signaled as those in NVIDIA GPUs. This might be explained by the different query methodologies used by
NVIDIA NVML, which has a wider time average frame of 1\,s. The MI250X utilization shows fast variations between 100 and 0 percentages highlighting the need for more robust energy tooling for further investigation. Energy consumption varies nearly linearly for all cases allowing for simple modeling of this benchmark.
On the H100, the overall integrated energy usage ($\approx$112--115\,kJ) shows little sensitivity, $\approx$2.6\% up to 1\,s, to the different resolutions. Importantly, the 1\,ms resolution yields false positive results, as shown in the spikes in Fig.~\ref{fig:resolution_qmcpack_1ms_h100}, which are a repeatable pattern in our experiments. Additionally, the 1\,s resolution is not enough to capture variations reinforcing the need for experimenting with finer resolutions.
On the A100, the power traces have more variability than for the H100. It is unclear if this is due to changes in the methodology for measuring power in more recent NVIDIA GPUs or if it is an actual hardware characteristic. Importantly, querying at 1\,ms introduces overhead in the time-to-solution on the A100, we ensured this is a repeatable pattern. Overall, both the A100 and H100 runs have similar energy characteristics, although as expected the H100 is more performance and energy efficient. As for AMD, energy characteristics are slightly different. As for the cause, either the time-to-solution (and therefore the energy consumption) is higher than it is on the A100 (and the H100).
Two things stand out: (i) $\leq$10\,ms measurements do not introduce noticeable overhead, but 1\,s resolutions might not be sufficient to capture variability for this workload. (ii) The DMC average power is lower for the AMD MI250X than for the NVIDIA GPUs at the expense of longer time-to-solution for a smaller number of walkers.
In all cases, temperature values remain nearly constant, thereby indicating that thermal solutions for cooling the devices run efficiently for these benchmarks.

\begin{figure}[tb!]
  \centering
  \subfloat[H100 10\,ms 68 walkers]{\includegraphics[width=0.5\textwidth,height=0.35\textwidth]{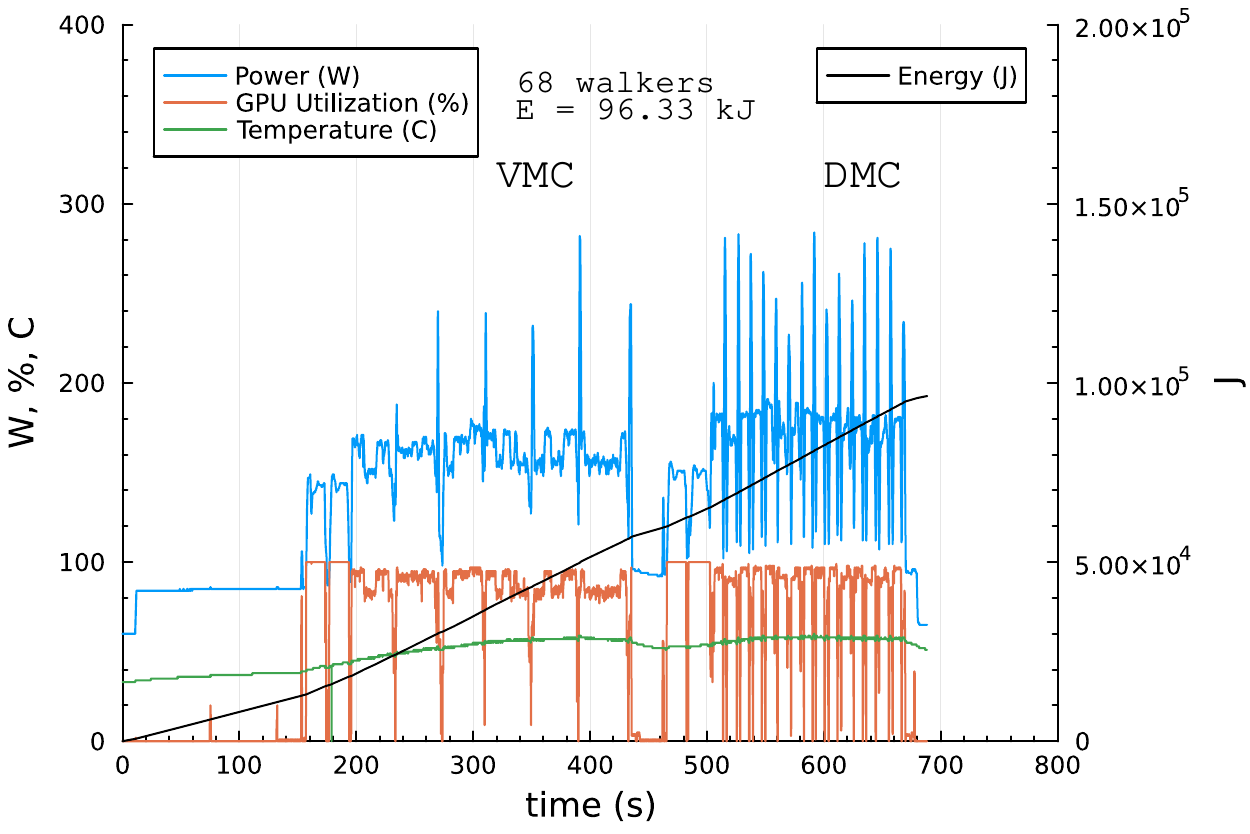}}%
  \subfloat[H100 10\,ms 100 walkers]{\includegraphics[width=0.5\textwidth,height=0.35\textwidth]{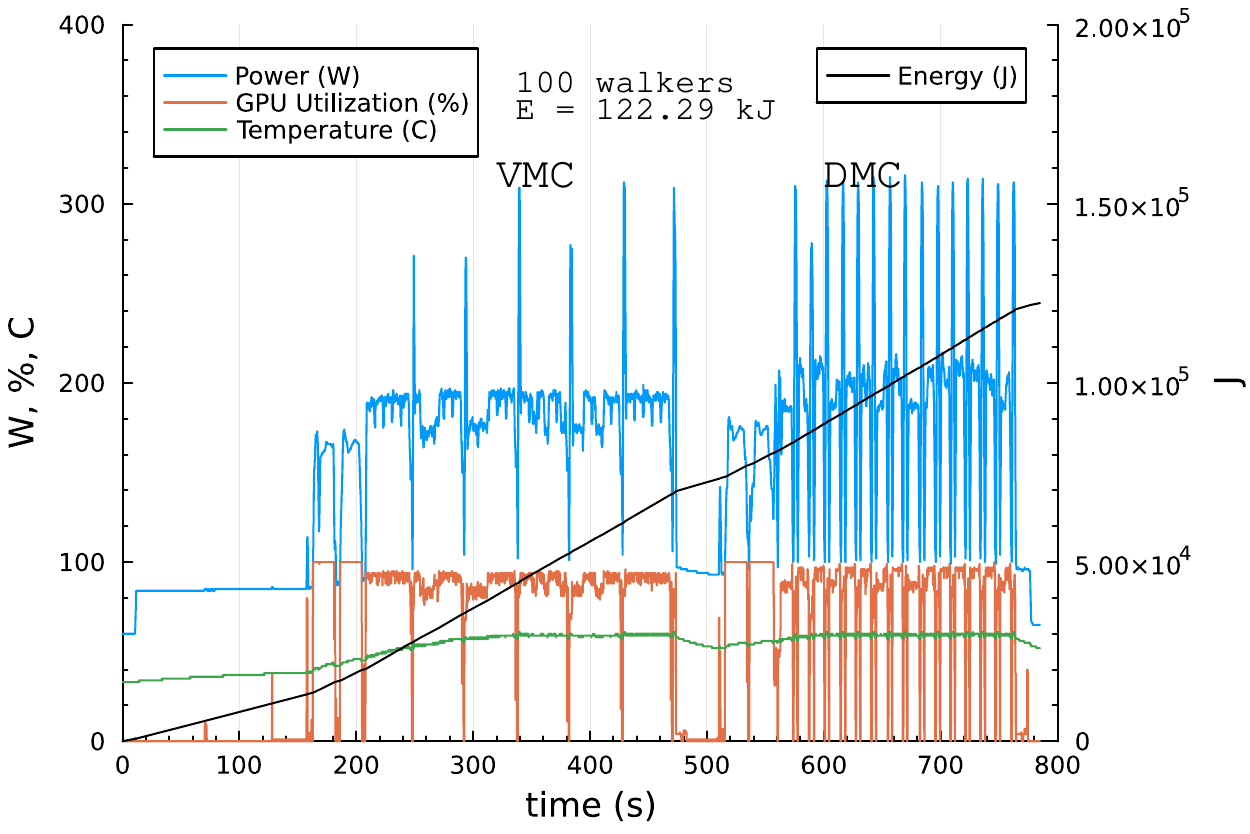}
  }

  \subfloat[A100 10\,ms 58 walkers]{\includegraphics[width=0.5\textwidth,height=0.35\textwidth]{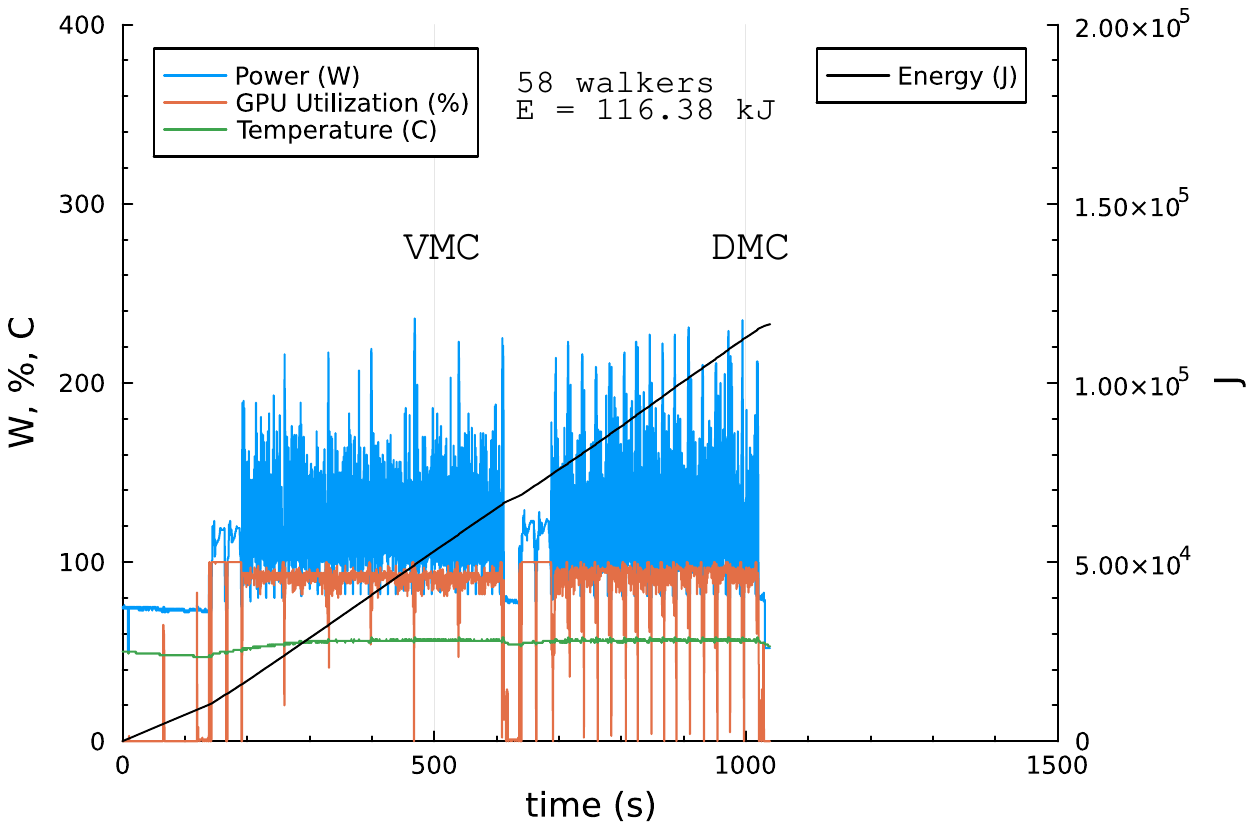}}%
  \subfloat[A100 10\,ms 84 walkers]{\includegraphics[width=0.5\textwidth,height=0.35\textwidth]{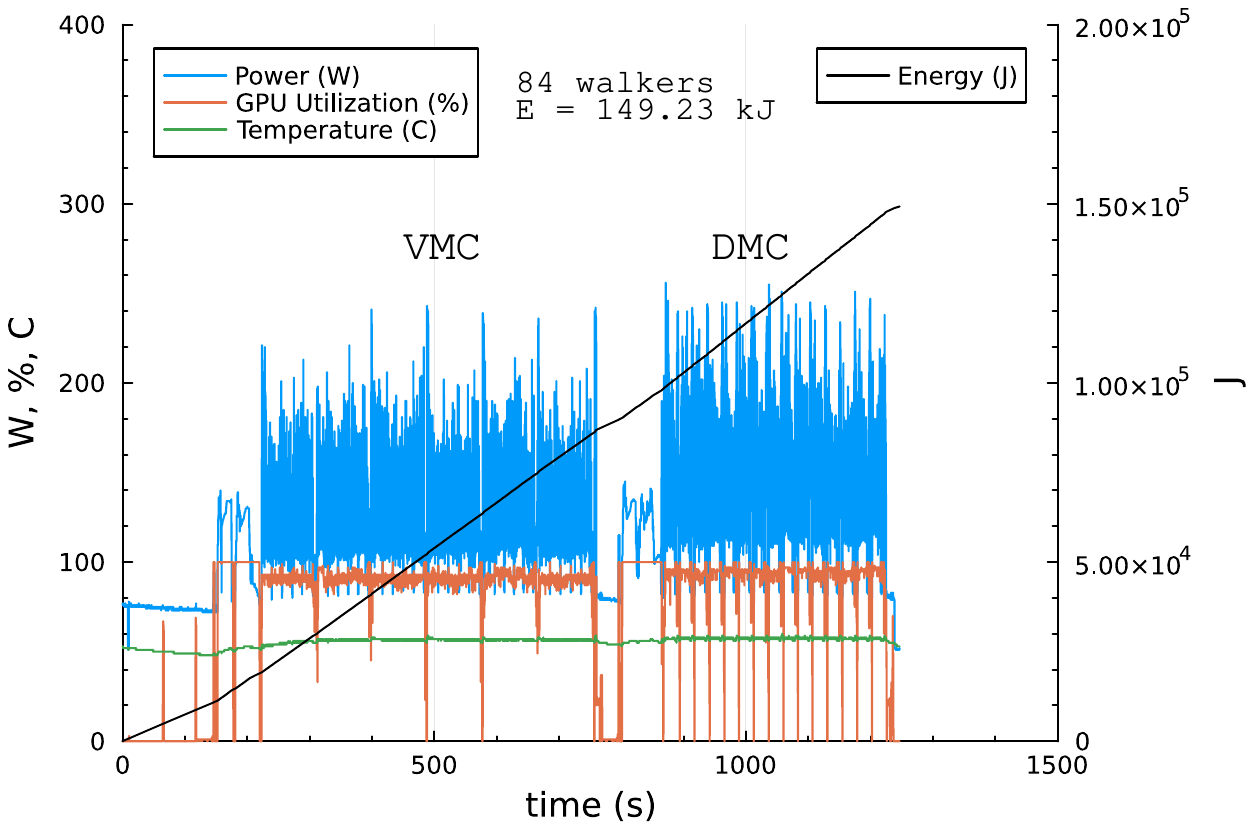}
  }
  \vspace{-2ex}
  \caption{Mixed-precision traces on NVIDIA H100 and A100 for (a) max double-precision walkers (68 and 58) and (b) max mixed-precision walkers (100 and 84).}
  \label{fig:mixed_qmcpack_nvidia}
  \vspace{-2ex}
\end{figure}

\begin{figure}[tb!]
\centering
\subfloat[MI250X 1\,ms 38 walkers]{\includegraphics[width=0.5\textwidth,height=0.35\textwidth]{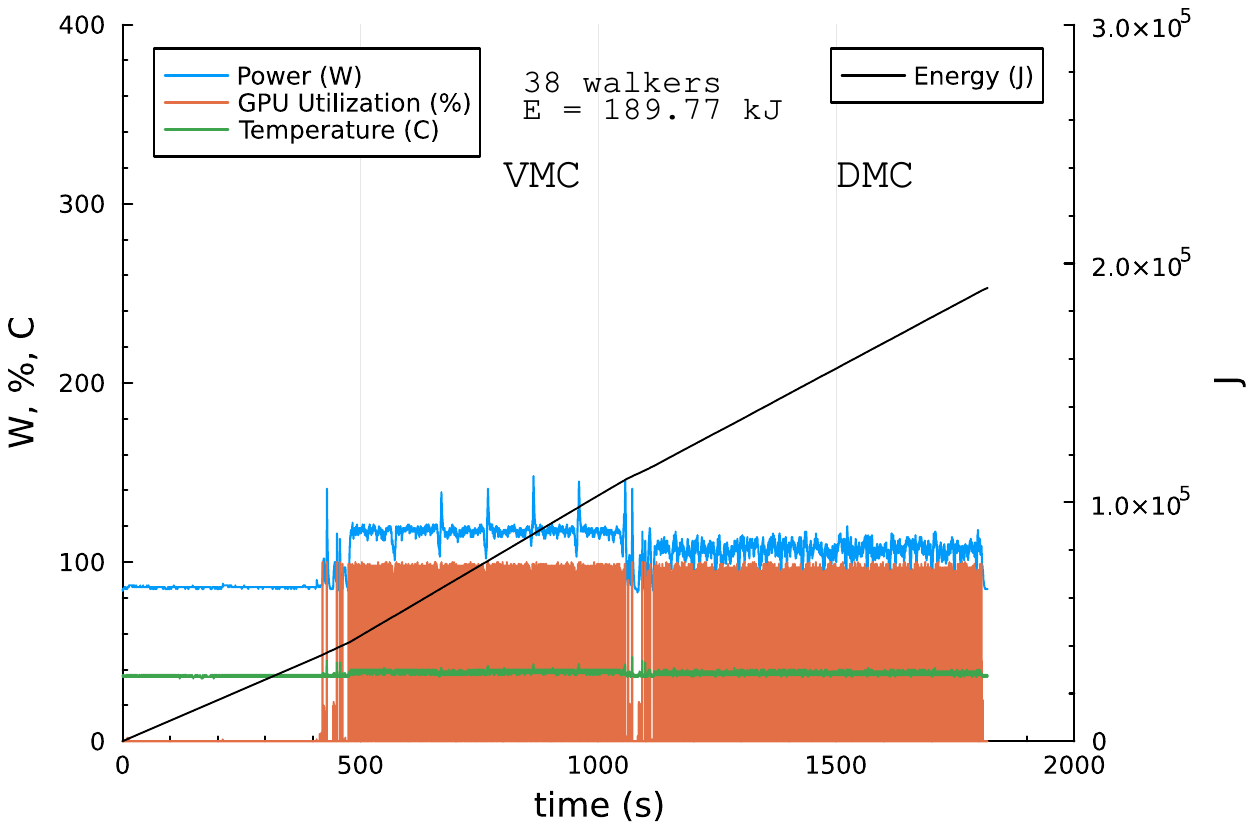}
}%
\subfloat[MI250X 1\,ms 52 walkers]{\includegraphics[width=0.5\textwidth,height=0.35\textwidth]{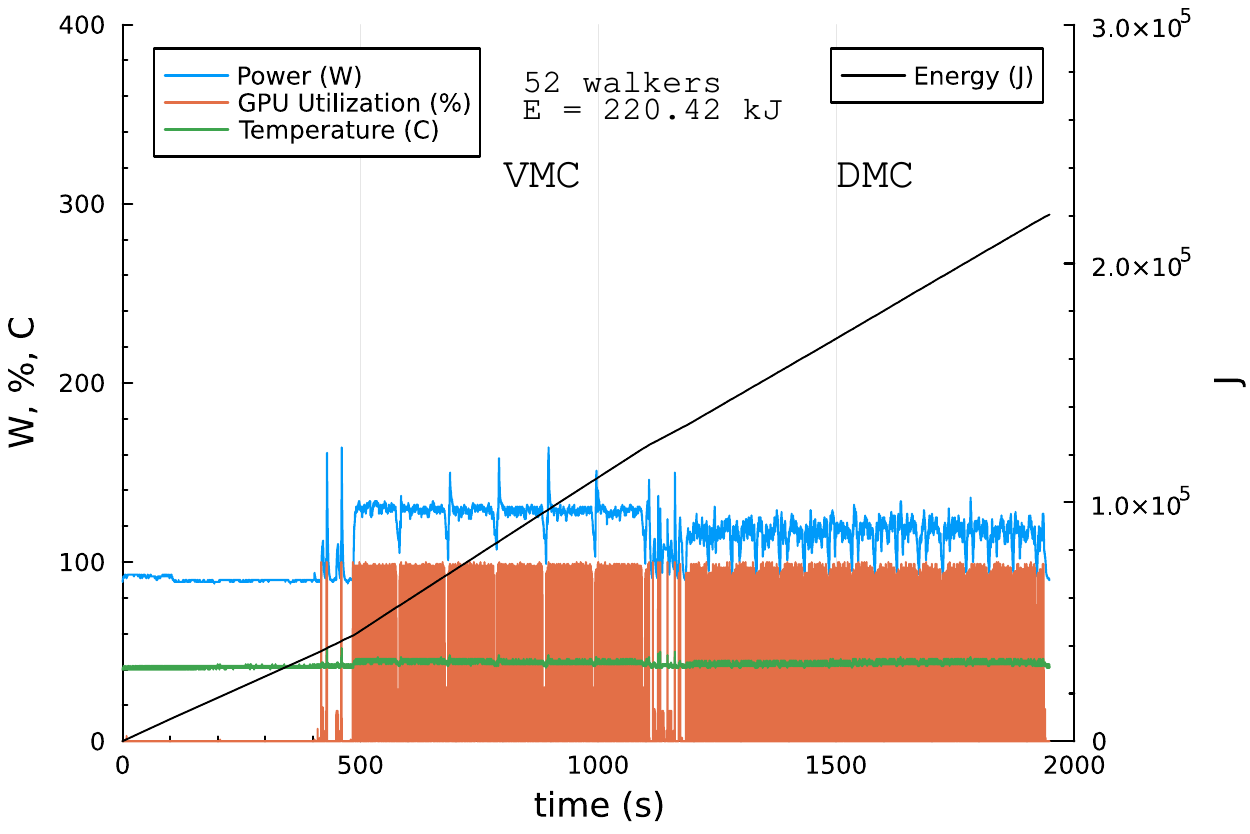}
}
\caption{AMD MI250X mixed-precision traces for (a) max double-precision walkers (38) and (b) max mixed-precision walkers (52).}
\vspace{-4ex}
\label{fig:mixed_qmcpack_amd}
\end{figure}

\paragraph{Mixed precision runs:} 
QMCPACK uses a mixed-precision (single/double) option that reduces memory consumption at the expense of less accurate results. This is applied in two ways: (i) a shorter time-to-solution for the same number of walkers, or (ii) more walkers
to fill the GPU's memory and increase utilization. These two options are illustrated in Figs.~\ref{fig:mixed_qmcpack_nvidia} and ~\ref{fig:mixed_qmcpack_amd} for the NVIDIA H100 and A100, and AMD MI250X GPUs, respectively. 

For these cases, (a) and (c) contain the same number of walkers as the maximum double-precision runs, and (b) contains the maximum number of walkers for a mixed-precision run.
In all cases, mixed-precision runs result in energy savings, thanks to faster times-to-solution than the double-precision runs.
This is more evident on the runs with NVIDIA's H100 and A100 GPUs than with AMD's MI250X GPU. Because some components keep double-precision representations, the savings are not expected to be close to 50\%.
QMCPACK's mixed-precision has not been thoroughly studied for the latest GPUs.
For example, this work identified different default behaviors (e.g., DMC mixed precision was running extra calculations) that has been corrected in QMCPACK.
Thus showing the importance of monitoring energy characteristics during development. A simple power and utilization trace can showcase hardware usage correlations with expected algorithm behavior.

\paragraph{Energy metric}
The DMC performance metric (throughput) is measured as the computational cost of advancing walkers at each step and block~\cite{10.1145/3581576.3581621}. We adapt this metric for energy efficiency (science/energy) by changing the denominator from time-to-solution to energy consumption in:

\begin{gather}
    \mathtt{Throughput_{Energy}} = \frac{\mathtt{walkers} \times \mathtt{blocks} \times \mathtt{steps}}{\mathtt{DMC\,energy}},
    \label{eqn:dmc_throughput_energy}
\end{gather}

\noindent
where $\mathtt{DMC\,energy}$ is the energy needed by the DMC region.

\begin{table}[tb!]
  \footnotesize
  \caption{Energy metrics for QMCPACK's DMC on the NiO a512 benchmark for multiple GPU configurations}
  \begin{center}
    \begin{tabular}{ |l|r|r|r|r| } 
      \hline 
      \textbf{Configuration} & \textbf{Walkers}  & \multicolumn{1}{c|}{\textbf{Throughput}} &  \textbf{Power} & \textbf{GPU}\\
               & \textbf{*Max}     &  \textbf{Energy (1/kJ)}\,\,\,     &  \textbf{(W)} &  \textbf{(\%)} \\
      \hline
      \hline
      \textbf{NVIDIA}     &     &        &         &  \\
      H100-mixed & *100 & 38.69  & 190.02  &  72.54\\
      H100-mixed &  *68 & 33.20  & 172.60  &  74.60 \\
      H100-double  &  *68 & 27.26  & 191.87  &  83.92 \\
      \hline
      H100-mixed &  84 &  37.56  &  182.31  &  74.69 \\
      H100-mixed &  58 &  32.42 &  174.24  &  75.93 \\
      H100-double  &  58 &  26.28 &  181.89  &  78.58 \\
      A100-mixed &  *84 & 25.25  & 136.10  &  85.76 \\
      A100-mixed &  *58 & 20.86  & 121.59  &  85.08 \\
      A100-double  &  *58 & 17.31  & 124.97  &  88.03 \\
      \hline
      A100-mixed &  52 & 21.41  & 122.22  &  83.84 \\  
      A100-mixed &  38 & 16.13  & 109.01  &  86.92 \\  
      A100-double  &  38 & 15.58  & 119.11  &  86.69 \\
      \hline
      \textbf{AMD} & & & & \\ 
      MI250X-mixed & *52  &  9.12    &  115.97  & 39.33  \\
      MI250X-mixed & *38  &  7.57    &  106.36  & 40.12  \\ 
      MI250X-double  & *38  &  6.19    &  112.61  & 39.27  \\
     \hline
    \end{tabular}
  \end{center}
  \label{tab:metric_qmcpack}
  \vspace{-2ex}
\end{table}

Metric results are presented in Table~\ref{tab:metric_qmcpack} along with the average GPU power and utilization. Using the maximum number of walkers for mixed precision on H100 leads to the greatest science/energy ratio, and running in double precision draws higher energy rates. Energy savings from mixed-precision for the same number of walkers are on the order or 6\%--25\%. We also added cases to compare H100 versus A100, and A100 versus MI250X runs for the same number of walkers. Energy improvements between A100 and H100 cases are roughly 1.5$\times$, whereas A100 runs are 2$\times$ more efficient than MI250X runs. Results for AMD's MI250X suggest that improvements are needed on either the software or vendor tool stack to achieve results comparable to NVIDIA's A100.

\subsection{AMReX-Castro Sedov}

\begin{figure}[tb!]
    \centering
    \includegraphics[width=0.6\textwidth,height=0.4\textwidth]{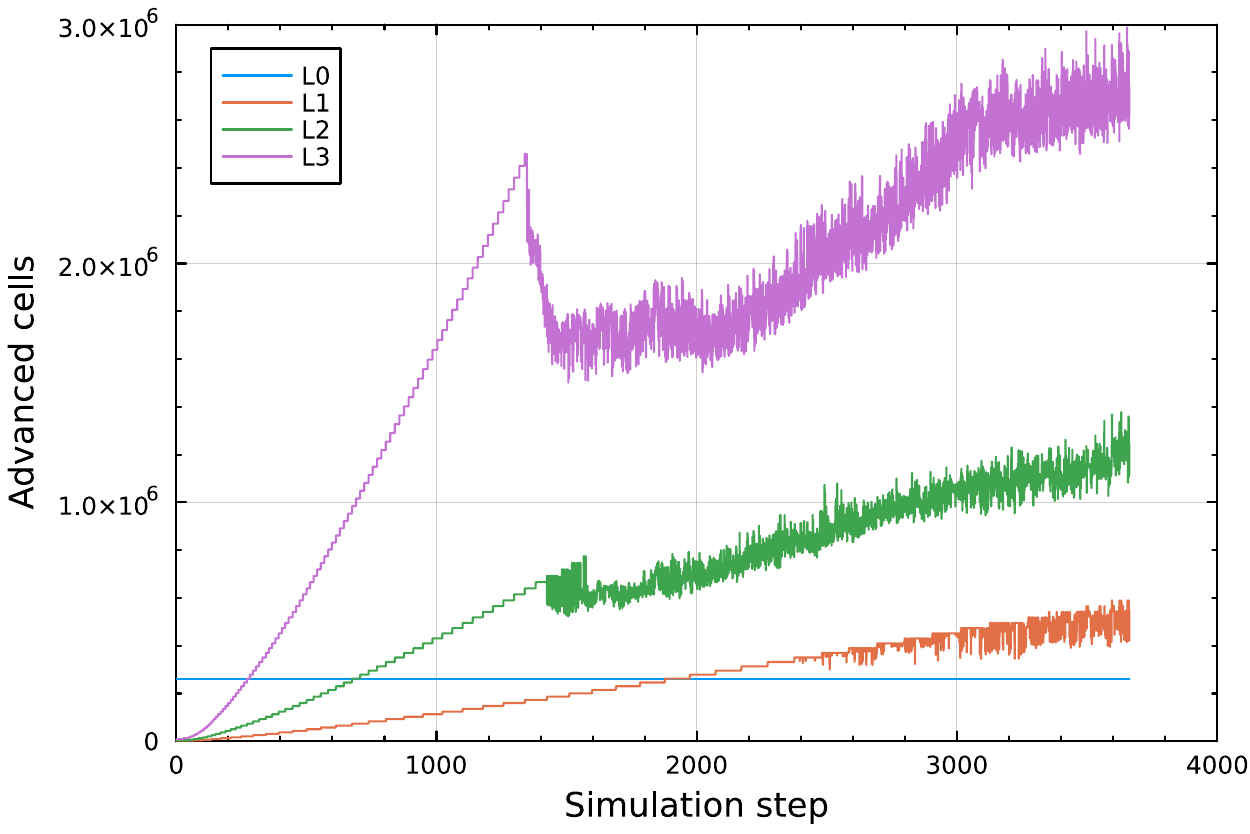}
    \caption{AMReX-Castro Sedov problem advanced cells per AMR mesh level as a function of simulation step.}
    \label{fig:trace_castro_cells}
\end{figure}

We ran the 2D Sedov case to capture an AMR simulation's influence on energy characteristics.
Compute activity comes from advancing a variable evolution (governed by PDEs) on cells placed at different mesh levels. Fig.~\ref{fig:trace_castro_cells} shows the evolution of four mesh levels (L0--L3) in a coarser-to-finer order as a function of simulation steps.
As seen, the finer mesh (L3) dominates the computation, and all of them except the base mesh (L0) evolve similarly. We expect that energy consumption would be dominated by this evolution and the use of double- or single-precision representations. 

\begin{figure}[tb!]
    \centering
    \subfloat[H100 10\,ms double]{\includegraphics[width=0.5\textwidth,height=0.35\textwidth]{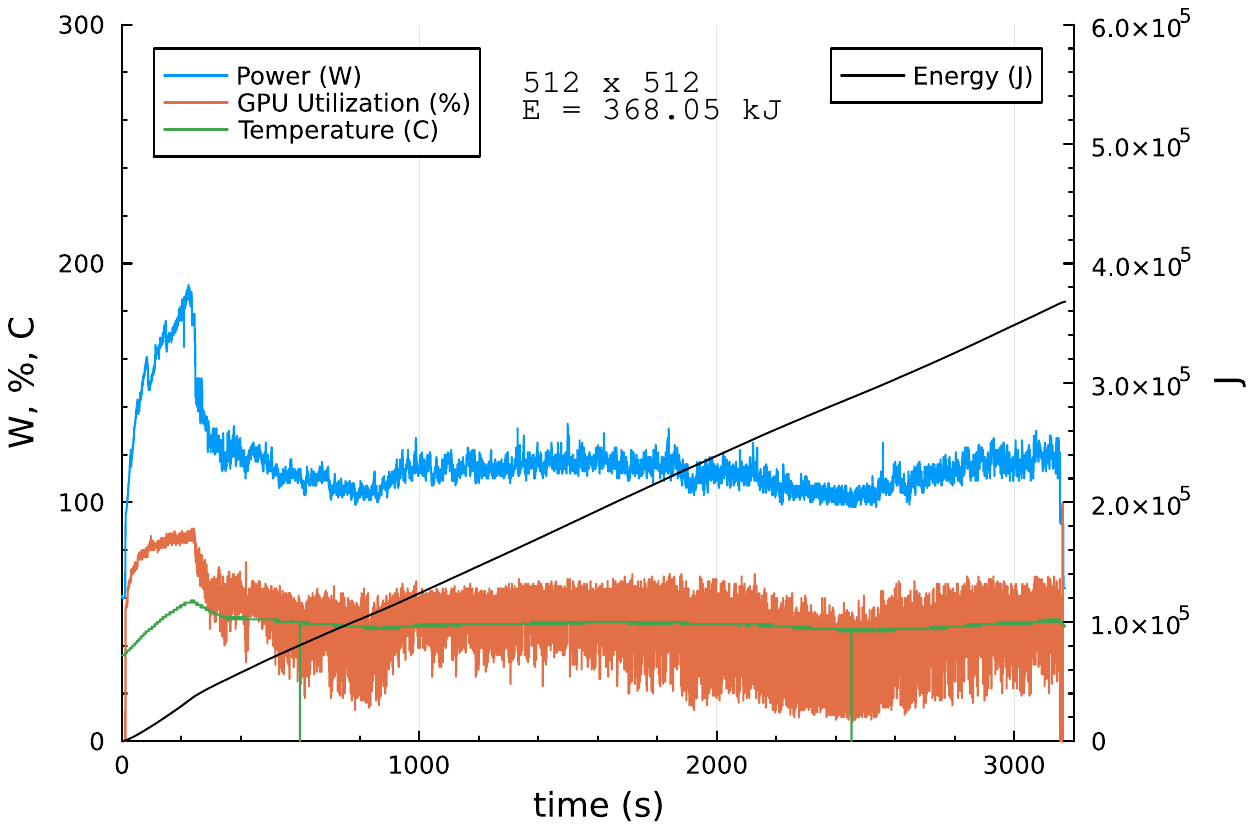}
    \label{fig:castro_sedov_h100_double}
    }%
    \subfloat[H100 10\,ms single]{\includegraphics[width=0.5\textwidth,height=0.35\textwidth]{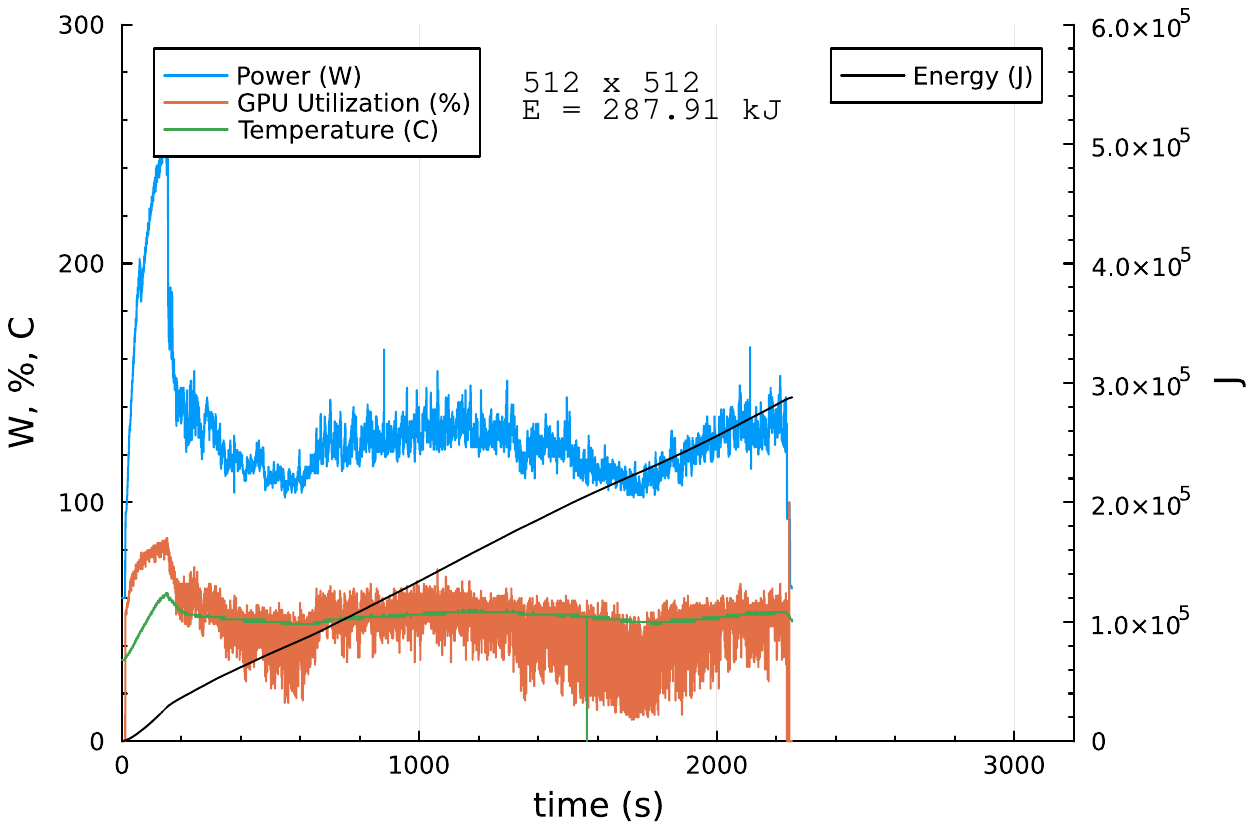}
    \label{fig:castro_sedov_h100_single}
    }
    
    \subfloat[A100 10\,ms double]{\includegraphics[width=0.5\textwidth,height=0.35\textwidth]{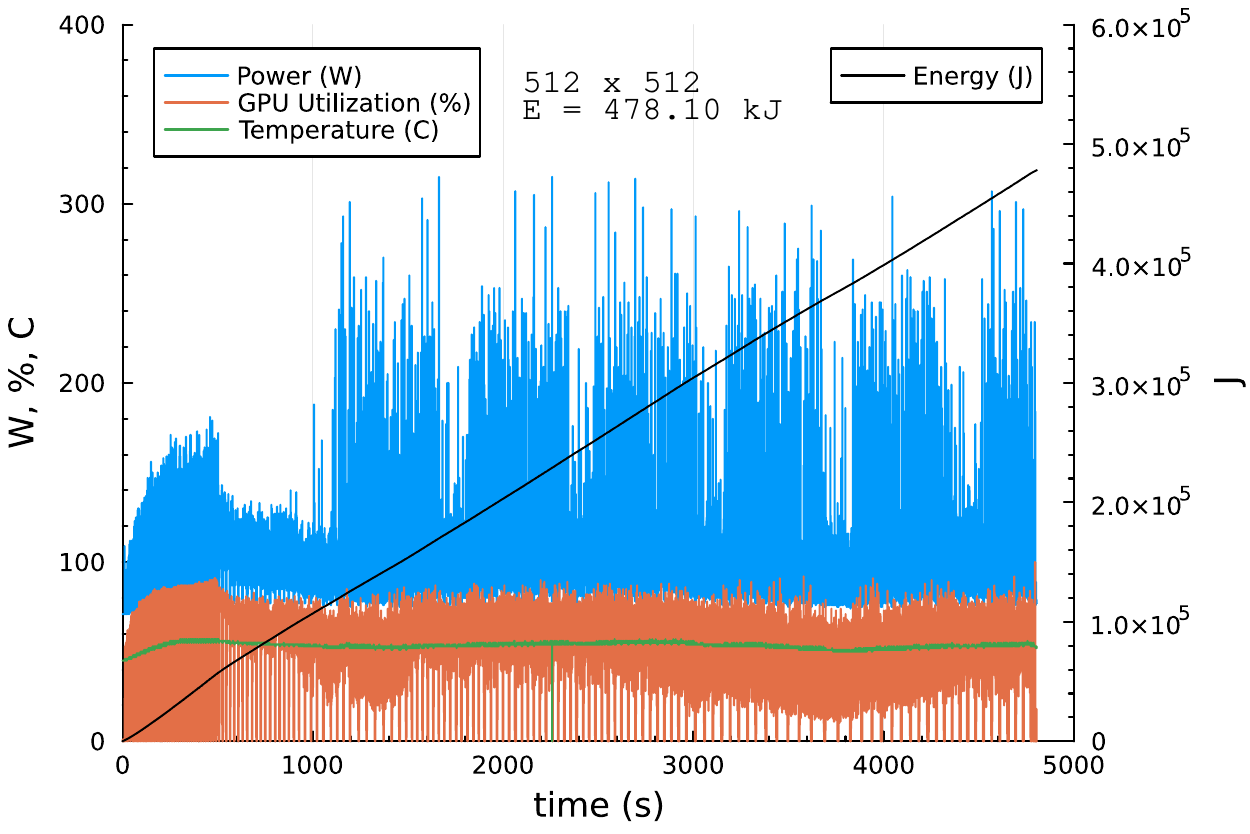}
    \label{fig:castro_sedov_a100_double}
    }%
    \subfloat[A100 10\,ms single]{\includegraphics[width=0.5\textwidth,height=0.35\textwidth]{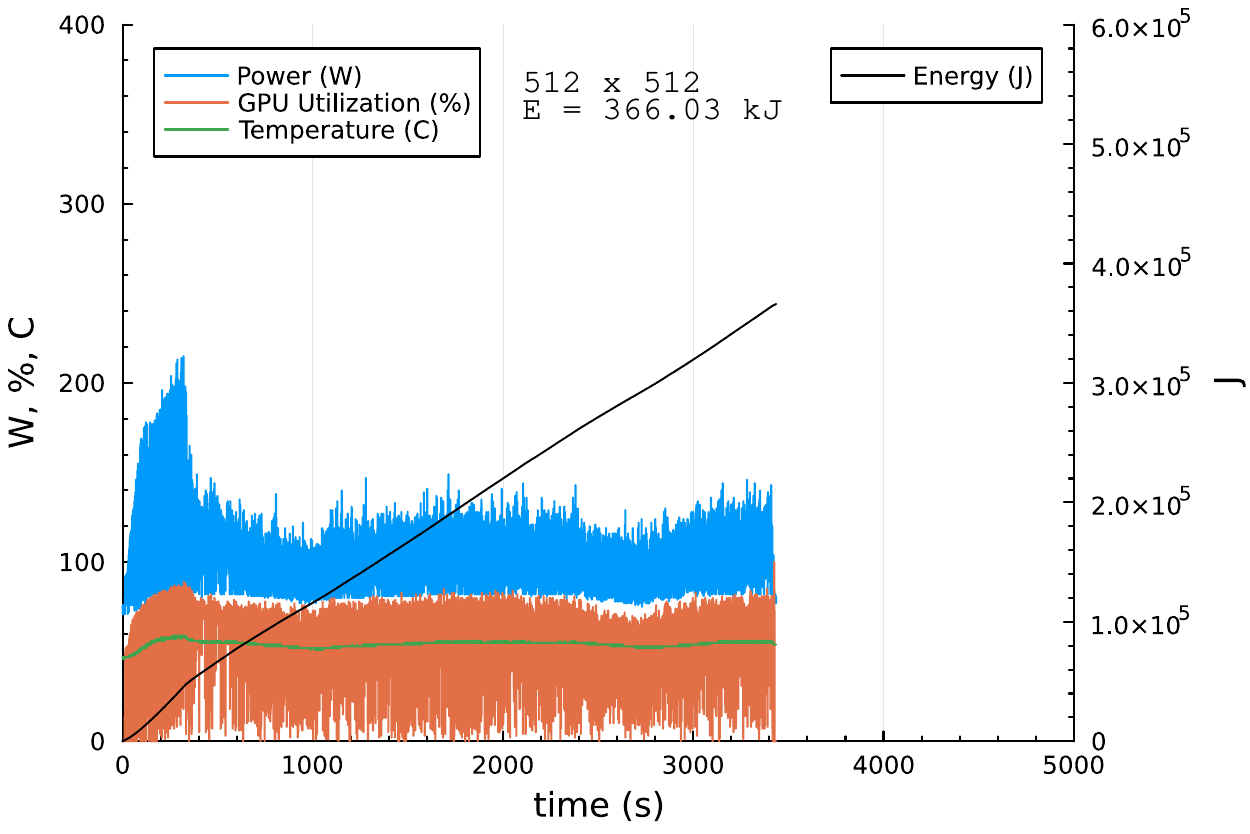}
    \label{fig:castro_sedov_a100_single}
    }
    
    \caption{AMReX-Castro Sedov energy characteristics on NVIDIA H100 and A100 for a 512~$\times$~512 base mesh with CFL = 0.25 using (a) double and (b) single precision.}
    \label{fig:castro_sedov_nvidia}
\end{figure}

\begin{figure}[tb!]
    \centering
    \includegraphics[width=0.5\textwidth,height=0.35\textwidth]{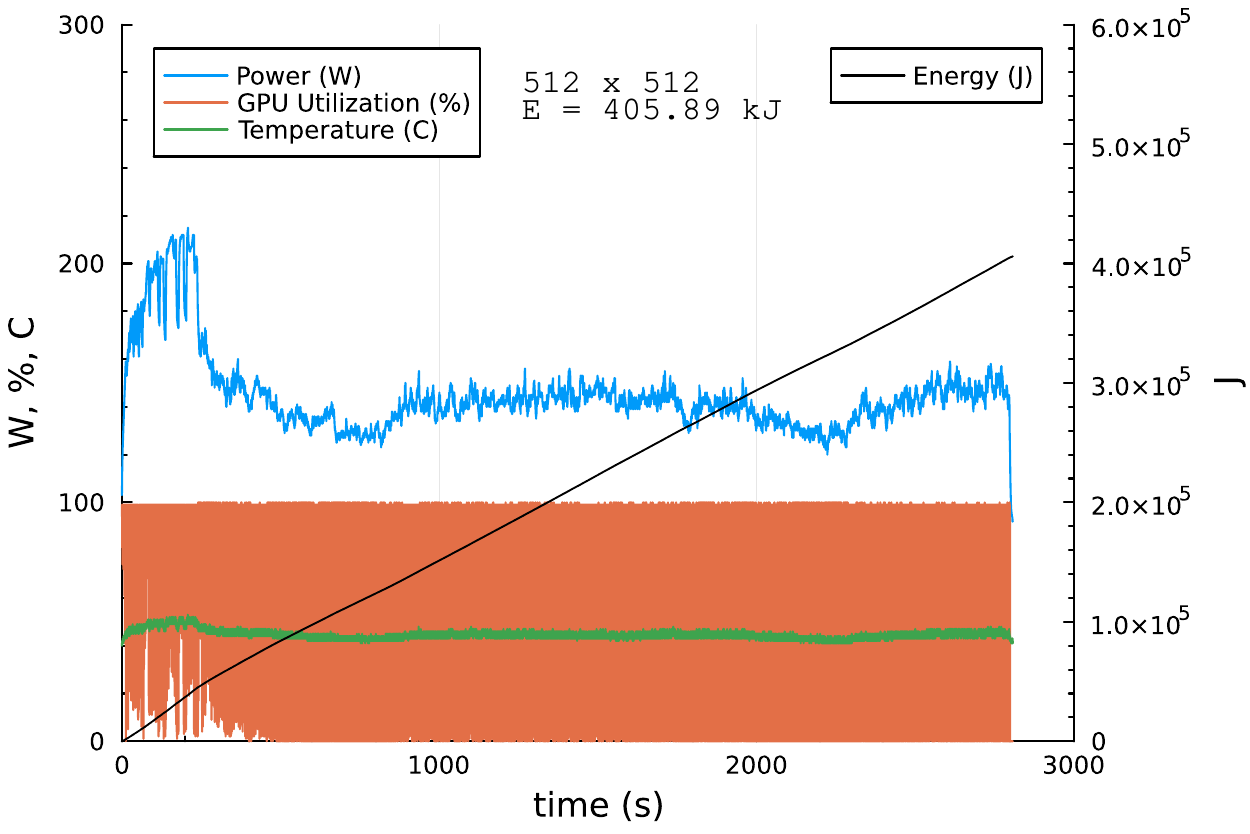}
    \caption{AMReX-Castro Sedov energy characteristics on an AMD MI250X for a 512~$\times$~512 base mesh using double precision, 1\,ms resolution, and CFL = 0.25.}
    \label{fig:castro_sedov_amd}
    \vspace{-2ex}
\end{figure}

\paragraph{Double/single-precision}
In all cases, we use the previously set query resolution for NVIDIA (10\,ms) and AMD (1\,ms) GPUs, and fixed the Courant-Friedrichs-Lewy (CFL) condition to a value of 0.25.  
Fig.~\ref{fig:castro_sedov_nvidia} shows the traces on NVIDIA H100 and A100 runs when using the largest possible base 512~$\times$~512 mesh and double- and single-precision representations.
As expected, power and utilization show a strong correlation with the finer mesh evolution in Fig.~\ref{fig:trace_castro_cells}. Power variability is high for the rapid evolution of the mesh levels (1,200--1,400 simulation steps), reaching an absolute peak and stabilizing until the end of the run. 
The use of single precision (Figs.~\ref{fig:castro_sedov_h100_single} and ~\ref{fig:castro_sedov_a100_single}) represents a faster code execution ($\approx$2200/3200 = 68\%) when compared with double precision (Figs.~\ref{fig:castro_sedov_h100_double} and ~\ref{fig:castro_sedov_a100_double}) but a larger fraction in terms of energy ($\approx$287/368 = 78\%). In fact, the initial peak power is higher than for double precision. Nevertheless, the H100 runs demonstrate the improvements in energy consumption over the previous A100 by using only 60\% and 78\% as much energy for the double- and single-precision cases, respectively.
For AMD, the MI250X is more energy and time efficient than the A100 for the Sedov case when using double precision (Fig.~\ref{fig:castro_sedov_amd}). Power characteristics are very similar to those observed on NVIDIA's H100.

We did not observe differences between double- and single-precision runs on the MI250X, so the latter is not provided.
Energy-efficiency metrics still need to be studied for this application as the influence of more complex performance and energy drivers (e.g., non-linear evolution in the AMR mesh sizes and CFL number) is highly problem- and hardware-dependent.

\section{Related Work}
\label{sec:Related-Work}

Energy efficiency in HPC has garnered significant attention in recent years at all levels (including applications, facilities, and tools).
Muriedas, et al. and Yang, et al. measured HPC application's energy consumption on Intel and NVIDIA GPUs~\cite{perun, YangAdamek2024-nvidiaSMI}.
Schieffer et al. characterized energy on AMD matrix cores using \texttt{rocm\_smi\_lib}~\cite{10590025}.
Other work has designed a variety of tools for predicting GPU power consumption.
This includes AccelWattch~\cite{KandiahPeverelle2021-accelWattch}, 
machine learning models to predict power consumption~\cite{lee2015-powertrain, shim2022-deeppm, WuGreathouse2015-gpuPowerModelingML}, and flexible simulator interfaces~\cite{SmithBruce2024-gem5Power}.
However, most of these approaches focus on modeling and simulation tools, unlike our work.
Simsek et al.~\cite{Simsek_2023,10820648} studied the energy efficiency of the SPH-EXA astrophysics application on CPUs and GPUs as well as the impact of dynamic frequency scaling using the open-source Power Measurement Toolkit~\cite{corda2022pmt}. Govind et
al.~\cite{govind2023comparing} investigated the power characteristics of scientific and machine learning applications on the
Perlmutter supercomputer.
Zhao et al. studied power traces of the popular Vienna Ab initio Simulation Package on NVIDIA A100 GPUs including power capping techniques~\cite{10820603}.
Foster et al.~\cite{10196597} studied the energy efficiency of machine learning benchmarks.
Mantovani et al. studied the performance and energy consumption of HPC workloads on Arm ThunderX2 CPU cores~\cite{mantovani2020performance}.
Mittal and Vetter presented a survey of methods for improving GPU energy efficiency~\cite{10.1145/2636342}.
Bridges et al. provided an overview of techniques for obtaining GPU power consumption data~\cite{ORNL-Bridges}.
At the facility level, Karimi et al. presented a system-wide HPC monitoring framework on the Summit supercomputer~\cite{10631049}.
Shin et al. introduced a comprehensive strategy for the sustainability of future HPC systems~\cite{10820562}.
Silva et al. presented a comprehensive survey on the energy of state-of-the-art supercomputers~\cite{silva2024review}.
Rrapaj et al. quantified the long-term energy consumption in systems of the National Energy Research Scientific Computing Center~\cite{10528943}.

\section{Discussion, Conclusions and Future Work}
\label{sec:Conclusions}

We quantified and analyzed the energy consumption characteristics of two science applications that are widely used in HPC systems---QMCPACK and AMReX-Castro---on recent
NVIDIA A100 and H100, and AMD MI250X GPUs. Kernel characteristics were mapped to the power, temperature, utilization, and energy traces
for different configurations, including reduced floating-point precision. Application-specific metrics were discussed compare the science-per-energy. Some key observations:

\begin{itemize}
    \item {\bf Observation 1:} small variability for the power, utilization, temperature traces and integrated energy calculations were observed for vendor APIs, NVML and rocm\_smi\_lib, in our application measurements.
    \item {\bf Observation 2:} reduced floating-point precision resulted in more energy-efficient runs, while not at an ideal 50\% rate these were on the order of 6\%--25\% on QMCPACK (on NVIDIA and AMD) and 45\% for AMReX-Castro (on NVIDIA only).
    \item {\bf Observation 3:} Energy-efficiency improvements (on the order of 1.5$\times$) were shown for NVIDIA's H100 over their A100. Room for improvement exists for AMD's GPU tools and applications as the ecosystem matures.
    \item {\bf Observation 4:} the proposed science-per-energy metric for QMCPACK allows for comparison across GPU vendors and generations. It factors in computational time aspects into the total energy required. A AMReX-Castro science-per-energy metric requires factoring-in the AMR variability in the number of cells advanced at every level.
\end{itemize}

  Next steps include expanding to more HPC-relevant scientific workloads, understanding energy at the different GPU component levels, and integrating modeling for design space exploration (e.g., GEM5~\cite{lowepower2020gem5simulatorversion200}). As HPC facilities energy costs increase, we conclude that this type of analysis at the application level is crucial in the co-design of future supercomputing architectures.

\section*{Acknowledgements}
This manuscript has been authored by UT-Battelle, LLC, under contract DE-AC05-00OR22725 with the US Department of Energy (DOE). The US government retains and the publisher, by accepting the article for publication, acknowledges that the US government retains a nonexclusive, paid-up, irrevocable, worldwide license to publish or reproduce the published form of this manuscript, or allow others to do so, for US government purposes. DOE will provide public access to these results of federally sponsored research in accordance with the DOE Public Access Plan (\url{https://www.energy.gov/doe-public-access-plan}).
This material is based on work supported by the DOE's Office of Science, Office of Advanced Scientific Computing Research through EXPRESS: 2023 Exploratory Research for Extreme Scale Science. PRCK was supported by the DOE's Office of Science, Basic Energy Sciences, Materials Sciences and Engineering Division as part of the Computational Materials Sciences Program and the Center for Predictive Simulation of Functional Materials.
This research used resources of the Oak Ridge Leadership Computing Facility and the Experimental Computing Laboratory at the Oak Ridge National Laboratory, which is supported by the DOE's Office of Science under Contract No. DE-AC05-00OR22725. WG would like to acknowledge Brandon Tran from the University of Wisconsin for the valuable discussion on NVML. 

\bibliographystyle{splncs04}
\bibliography{references}

\end{document}